# The aug-cc-pVnZ-F12 Basis Set Family: Correlation Consistent Basis Sets for Explicitly Correlated Benchmark Calculations on Anions and Noncovalent Complexes

*Nitai Sylvetsky, Manoj K. Kesharwani, and Jan M.L. Martin\**

Department of Organic Chemistry, Weizmann Institute of Science, 76100 Reḥovot, Israel.

Email: gershom@weizmann.ac.il. FAX: +972 8 934 3029

**Abstract**

We have developed a new basis set family, denoted aug-cc-pVnZ-F12 (or aVnZ-F12 for short), for explicitly correlated calculations. The sets included in this family were constructed by supplementing the corresponding cc-pVnZ-F12 sets with additional diffuse functions on the higher angular momenta (i.e., additional d-h functions on non-hydrogen atoms, and p-g on hydrogen), optimized for the MP2-F12 energy of the relevant atomic anions. The new basis sets have been benchmarked against electron affinities of the first- and second-row atoms, the W4-17 dataset of total atomization energies, the S66 dataset of noncovalent interactions, the BEGDB water clusters subset, and the WATER23 subset of the GMTKN24 and GMTKN30 benchmark suites. The aVnZ-F12 basis sets displayed excellent performance, not just for electron affinities but also for noncovalent interaction energies of neutral and anionic species. Appropriate CABS (complementary auxiliary basis sets) were explored for the S66 noncovalent interactions benchmark: between similar-sized basis sets, CABS were found to be more transferable than generally assumed.

**Introduction**

Wavefunction *ab initio* calculations are, inherently, a two-dimensional convergence problem: for the most part, getting 'the right answer for the right reason' entails achieving convergence in terms of both (1-particle) basis set size and the level of (n-particle) electron correlation treatment.[1] For systems well described by a single reference determinant (i.e., excited states are energetically well-separated from the ground state), the "gold standard" CCSD(T)[2,3] (coupled cluster[4–7] with all single and double substitutions[8] and quasiperturbative triples[2,3]) method is known to produce results rather close to the n-particle limit (i.e., full configuration interaction or FCI) at a tiny fraction of the latter's cost. This leaves basis set convergence as the main bottleneck.

In terms of conventional one-particle basis sets, basis set convergence is rather slow.[9–11] Nevertheless, its monotonic character enables the use of extrapolation techniques when appropriately systematic basis set sequences are used.[12–15] The correlation consistent basis set family (denoted cc-pVnZ), developed by Dunning et al.,[16–18] has gained great popularity for displaying smooth, systematic convergence of energies and other calculated properties toward the complete basis set (CBS) limit as the cardinal number (*n* in cc-pVnZ) is increased. This feature permits the systematic improvement of *ab initio* calculations toward the exact solution of the time-independent, nonrelativistic, electronic Schrödinger equation − by systematically improving not only the electron correlation method (for example, employing CCSD(T) instead of CCSD), but also the basis set (i.e., cc-pVTZ instead of cc-pVDZ). Thus, it also paved the way for a large number of extrapolation formulas that can be used to estimate the CBS limit (see, e.g., Refs.[15,19]).

In order to obtain accurate electron affinities (EAs), a balanced description of the neutral and (spatially much more extended) anionic wavefunctions requires the inclusion of "diffuse" (small-exponent) basis functions for describing the long-range region. Kendall et al.[20] augmented their cc-pVnZ basis sets with one additional diffuse function for each angular momentum symmetry present within that basis. Specifically, additional s and p functions were optimized for the appropriate atomic anions to minimize the SCF energy, while functions with higher angular momenta (d,f,g,… etc., which do not contribute to the SCF energy in a spherical system) were optimized for the CISD correlation energy of the anions. As expected, the resulting "augmented" basis sets, denoted aug-cc-pVnZ (or, sometimes, AVnZ for short), were found to yield substantially better EAs than their unaugmented counterparts.[20] It had also been realized (e.g.,[21]) that diffuse functions accelerate basis set convergence for non-covalent interactions (NCIs).[22,23,24]

Aside from basis set extrapolation techniques, basis set convergence can be greatly accelerated by means of explicitly correlated (F12) methods,[25–29] in which some "geminal" terms that explicitly depend on the interelectronic distance are added to the 1-particle basis set.[30–37] Unfortunately, it has been shown that combining F12 methods with basis sets not specifically optimized for them may lead to erratic, non-monotonic convergence to the basis set limit.[38–41] In order to overcome this difficulty, the cc-pVnZ-F12 basis set family (or VnZ-F12 for short, consisting of F12 counterparts of the conventional cc-pVnZ sets) was developed by Peterson et al.;[42–44] these basis sets, which were optimized at the MP2-F12 level in the presence of the appropriate geminal terms, appear to exhibit smoother convergence behavior in F12 calculations than orbital-optimized cc-pVnZ basis.[45] After displaying superior performance for the W4-15 atomization energies benchmark,[41] CCSD-F12/VnZ-F12 calculations were recently adopted in the Wn composite protocols developed by our group, as they substantially reduce computational cost (manifested in CPU time and especially memory and scratch disk requirements) without sacrificing accuracy.[41]

The VnZ-F12 basis sets contain diffuse functions for s and p orbitals, but not for the higher angular momenta. As a result, in a recent benchmark study[46] carried out by our group on water clusters $(H_2O)_{n=2-10,20}$, $(H_3O^+)(H_2O)_{n=1-6}$, and $(OH^-)(H_2O)_{n=1-6}$, atypically poor basis set convergence was found for the anionic species.[46] In light of the above discussion regarding the aug-cc-pVnZ basis sets, one might suspect that this issue is much more general than the solvated hydroxides case, and expect it to affect calculations of EAs and NCIs as well as calculations involving any anions whatsoever. If that is so, augmenting the VnZ-F12 basis sets with additional diffuse functions, in a strategy similar to the one embraced in the aug-cc-pVnZ development procedure, will help recover the smooth convergence behavior of the correlation consistent basis set family for such systems. In the present paper, we will develop and present such augmented basis sets for explicitly correlated calculations, to be denoted aug-cc-pVnZ-F12 (or aVnZ-F12, for short). By in-depth benchmarking, we will also assess their performance with respect to their non-augmented counterparts for calculating EAs, total atomization energies, neutral NCIs, and anionic NCIs.

**Computational Details**

Most conventional and explicitly correlated *ab initio* calculations were carried out using the MOLPRO 2015.1 program system[47,48] running on the Faculty of Chemistry cluster at the Weizmann Institute of Science. Some additional post-CCSD(T) calculations for the electron affinities are discussed in that section.

Explicitly correlated RI-MP2 and coupled-cluster single point energy calculations were performed using the cc-pVnZ-F12 basis sets (n = D, T, Q, 5),[43,44] in addition to our own proposed aug-cc-pVnZ-F12 sets (their development process will be discussed in the 'Basis set optimization' subsection below). As prescribed in Ref. [38], the geminal exponent ($\beta$) values were set to 0.9 for cc-pVDZ-F12 and 1.0 for both cc-pVTZ-F12 and cc-pVQZ-F12 basis sets. For the cc-pV5Z-F12 basis set, $\beta=1.2$ was used, as recommended in Ref. [45]. In addition, "CABS corrections"[36,49] were employed throughout for the calculated SCF components. The 3C(Fix) ansatz was used throughout.[33]

Explicitly correlated MP2-F12 or CCSD(F12*) calculations require, in practice, three auxiliary basis sets (which we will denote by their acronyms in the MOLPRO program system[47]) in order to circumvent the need for three-and four- electron integrals:

- a "JKfit" basis set for the expansion of Coulomb (J) and exchange (K) type integrals, such as also used in density fitting Hartree-Fock or hybrid DFT calculations. In the present work, those were taken from Ref.[50];

- an "MP2fit" basis set for the RI-MP2 approximation, such as also used in density fitting MP2 (a.k.a. resolution of the identity MP2, RI-MP2) or double-hybrid DFT calculations. See Refs. [51,52] for the MP2fit basis sets used in the present work;

- an "OptRI" or "CABS" (complementary auxiliary basis set[53]) used for the evaluation of the F12-specific matrix elements. This aspect is discussed in detail in the Appendix.

The primary explicitly correlated CCSD approximation considered in this work is CCSD-F12b[54]. In a previous study (Ref.[45]; see also Ref.[55] for harmonic frequencies), we considered CCSD(F12*) (also known as CCSD-F12c)[54] instead of CCSD-F12b and found that the difference between the two approaches is only significant for the cc-pVDZ-F12 basis set, which is manifestly inadequate for molecular atomization energies but is sometimes adequate for EAs and NCIs. Thus, CCSD(F12*)/cc-pVDZ-F12 will also be considered where relevant.

For conventional *ab initio* calculations we used Dunning correlation consistent cc-pVnZ (n = D, T, Q, 5) basis sets and their diffuse-function augmented counterparts aug-cc-pVnZ.[16,20,56,57] Similar to our previous work[58] and long-standing practice, we combined regular cc-pVnZ basis sets on hydrogen with the corresponding diffuse-function augmented basis sets on all other atoms. We denote this as haVnZ for short. (This practice is also variously known in the literature as aug'-cc-pVnZ or A'VnZ for short,[59] or jul-cc-pVnZ.[60]) In addition, the ACVnZ basis sets by Dunning and coworkers,[61,62] ano-pVnZ+ and sano-pVnZ basis sets by Valeev & Neese,[63] and the awCVnZ and d-aug-cc-pwCVnZ sets by Peterson and coworkers[62] were also considered in parts of our work.

Basis set extrapolations were carried out using the usual two-point formula:

$$E_\infty = E(L) - [E(L) - E(L-1)] / \left[\left(\frac{L}{L-1}\right)^\alpha - 1\right] \quad (1)$$

where L is the angular momentum in the basis set and α an exponent specific to the level of theory and basis set pair. In the present study, basis set extrapolation exponents (α) were taken from the compilation in Table 2 of Ref[58]. For CCSD-F12(*)/cc-pV{D,T}Z, the extrapolation exponent 3.0598 was obtained by following the procedure described in Ref. [38]

The interaction energies of water clusters have been investigated with and without Boys–Bernardi[64] counterpoise corrections. The counterpoise-corrected interaction energy of the dimer AB is defined as:

$$D_{cp} = E[AB] - E[A(B)] - E[B(A)] \quad (2)$$

while the uncorrected ("raw") dissociation energy is:

$$D_{raw} = E[AB] - E[A] - E[B] \quad (3)$$

In these equations, E[A(B)] represents the total energy of monomer A in the presence of the basis functions on monomer B, and conversely for E[B(A)]. As $D_{cp}$ tends to converge to the basis set limit from the underbinding direction, and $D_{raw}$ from the overbinding direction, the average:

$$D_{half-half} = (D_{cp} + D_{raw})/2 \quad (4)$$

(i.e., "half-counterpoise"[65]) suggests itself as an alternative that exhibits more rapid basis set convergence in both conventional[65] and explicitly correlated[66] calculations.

The generalization of the counterpoise method to three- and more-body systems is not unique:[67] the most commonly employed generalization, which we shall also apply in the present work, appears to be the site-site function counterpoise method of Wells and Wilson,[68] in which, for a trimer:

$$D_{cp} = E[ABC] - E[A(B)(C)] - E[(A)B(C)] - E[(A)(B)C] \qquad (5)$$

and similarly for tetramers and larger.

**Results and Discussion**

**Basis set optimization**

As noted above, the VnZ-F12 basis sets of Peterson *et al.* contain only s and p diffuse functions for non-hydrogen atoms.[43–45] Following a strategy similar to that in Ref.[20], we augmented the VnZ-F12 basis sets with diffuse functions on the higher angular momenta: i.e., we added a single d function (or p, for hydrogen) to the VDZ-F12 basis set, 1d1f functions (1p1d for hydrogen) to VTZ-F12, 1d1f1g functions (1p1d1f for hydrogen) to VQZ-F12, and so forth. The additional diffuse function exponents were obtained by minimizing the MP2-F12 energy for the atomic anions, and are presented in Tables 1-2 for the elements B–Ar. For optimizing these aVnZ-F12 (n=D-5) sets, we employed an auxiliary basis set combination that ought to be saturated, namely, awCV5Z/MP2fit, aV5Z+d/JKfit, and aV5Z+d/OptRI.

The $N^-$, $Ne^-$, and $Ar^-$ atomic anions are unbound, and thus direct optimization of diffuse functions for these anions is not possible. Hence, we avail ourselves of the same expedient as Kendall et al.[20] in their development of the aug-cc-pVnZ basis sets: the diffuse function exponents for N at each angular momentum were determined by cubic interpolation (with the atomic number Z as the independent variable) between values for B, C, O, and F, while those for Ne and Ar were obtained through cubic and quartic extrapolation, respectively.

**TABLE 1.** Optimized exponents (ζ) for the diffuse functions in the aVDZ-F12 and aVTZ-F12 basis sets for elements B-Ar. The number of basis functions for both VnZ-F12 and aVnZ-F12 sets is also given.

| Element | Nbas(VDZ-F12) | Nbas(aVDZ-F12) | ζ(d) | Nbas(VTZ-F12) | Nbas(aVTZ-F12) | ζ(d) | ζ(f) |
|---|---|---|---|---|---|---|---|
| H[a] | 9 | 13 | 0.0743 | 18 | 27 | 0.0594 | 0.1065 |
| B |  |  | 0.0730 |  |  | 0.0637 | 0.0859 |
| C |  |  | 0.0968 |  |  | 0.0916 | 0.1255 |
| N | 30 | 35 | 0.1242 | 53 | 65 | 0.1138 | 0.1485 |
| O |  |  | 0.1562 |  |  | 0.1354 | 0.1935 |
| F |  |  | 0.1940 |  |  | 0.1614 | 0.2986 |
| Ne |  |  | 0.2385 |  |  | 0.1969 | 0.3950 |
| Al |  |  | 0.0436 |  |  | 0.0345 | 0.0567 |
| Si |  |  | 0.0636 |  |  | 0.0512 | 0.0840 |
| P | 39 | 44 | 0.0762 | 62 | 74 | 0.0655 | 0.0961 |
| S |  |  | 0.0920 |  |  | 0.0821 | 0.1146 |
| Cl |  |  | 0.1140 |  |  | 0.0947 | 0.1446 |
| Ar |  |  | 0.1377 |  |  | 0.1059 | 0.1748 |

[a]For hydrogen, we present the optimized *p* and *d* exponents (instead of *d* and *f* exponents).

**TABLE 2.** Optimized exponents (ζ) for the diffuse functions in the aVQZ-F12 and aV5Z-F12 basis sets for elements B-Ar. The number of basis functions for both VnZ-F12 and aVnZ-F12 sets is also given.

| Element | Nbas(VQZ-F12) | Nbas(aVQZ-F12) | ζ(d) | ζ(f) | ζ(g) | Nbas(V5Z-F12) | Nbas(aV5Z-F12) | ζ(d) | ζ(f) | ζ(g) | ζ(h) |
|---|---|---|---|---|---|---|---|---|---|---|---|
| H[a] | 34 | 50 | 0.0553 | 0.0911 | 0.1053 | 80 | 105 | 0.0538 | 0.0818 | 0.1020 | 0.1875 |
| B |  |  | 0.0505 | 0.0628 | 0.1264 |  |  | 0.0445 | 0.0646 | 0.0806 | 0.0955 |
| C |  |  | 0.0758 | 0.0974 | 0.1972 |  |  | 0.0635 | 0.0955 | 0.1209 | 0.1114 |
| N |  |  | 0.1062 | 0.1316 | 0.2101 |  |  | 0.0786 | 0.1260 | 0.1672 | 0.1694 |
| O | 87 | 108 | 0.1318 | 0.1711 | 0.2323 | 134 | 166 | 0.0980 | 0.1608 | 0.2231 | 0.2409 |
| F |  |  | 0.1429 | 0.2216 | 0.3309 |  |  | 0.1299 | 0.2407 | 0.2869 | 0.2970 |
| Ne |  |  | 0.1635 | 0.2729 | 0.3852 |  |  | 0.1543 | 0.2624 | 0.3568 | 0.3895 |
| Al |  |  | 0.0334 | 0.0450 | 0.0617 |  |  | 0.0252 | 0.0388 | 0.0495 | 0.0826 |
| Si |  |  | 0.0463 | 0.0682 | 0.0973 |  |  | 0.0391 | 0.0588 | 0.0714 | 0.0869 |
| P |  |  | 0.0595 | 0.0808 | 0.1094 |  |  | 0.0481 | 0.0676 | 0.0868 | 0.1262 |
| S | 96 | 117 | 0.0718 | 0.1137 | 0.1408 | 140 | 172 | 0.0548 | 0.0873 | 0.1110 | 0.1481 |
| Cl |  |  | 0.0851 | 0.1417 | 0.1784 |  |  | 0.0749 | 0.1164 | 0.1413 | 0.1850 |
| Ar |  |  | 0.0991 | 0.1847 | 0.2473 |  |  | 0.0927 | 0.1298 | 0.1896 | 0.2016 |

[a]For hydrogen, we present the optimized *p, d, f,* and *g* exponents (instead of *d, f, g,* and *h* exponents).

In order to check their validity, a few 'checks and balances' can be applied to the optimized exponents:

1. within a row, the geometric mean of the exponents should increase roughly as $(Z^*)^2$, where $Z^*$ is the 'shielded' nuclear charge (first Ahlrichs-Taylor rule, see Ref.[69]);

2. exponents in a larger set (e.g., 3d) should mesh with the next smaller set (e.g., 2d);

3. the ratio of successive exponents in a set should be about two or higher, and all gaps within the core or valence parts should roughly be of the same order;[70]

4. the geometric mean of the exponents for each angular momentum should roughly increase by a factor of 1.2 for each step up in L (second Ahlrichs-Taylor rule[69]).

A number of the diffuse exponents for the noble gases obtained by the extrapolation mentioned above violate the above conditions because of polynomial oscillation: in such cases, we reduced the order of the polynomial by one and carried out least squares fitting, and if this still did not help, resorted to Ruedenberg even-tempered expansion from the other exponents of that angular momentum. Details are given in the ESI.

**Electron Affinities**

For atomic systems, electron affinities (EAs) are defined as -

$$EA = E(neutral) - E(anion) \qquad (6)$$

For first- and second-row atoms, EAs have been very accurately measured. Thence, before comparing the performance of the VnZ-F12 and aVnZ-F12 basis set families for calculating EAs, we shall briefly discuss the agreement between our best calculated results and up-to-date experimental values found in the literature.[71–79]

The CCSD(F12*) component was computed at the CCSD(F12*)/d-aug-cc-pwCV5Z level, while the (T) component was computed conventionally at the CCSD(T)/aug-cc-pCV{5,6}Z level (ACV{5,6}Z for short), where the {n-1,n} notation indicates extrapolation from that pair of basis sets. Post-CCSD(T) electron correlation corrections were obtained using the string-driven general coupled cluster code of Kállay and coworkers[80–83] as implemented in their MRCC program package.[84] The full CCSDT–CCSD(T) difference and the contribution of quasiperturbative connected quadruple excitations[81–83] (Q) were obtained with the aug-cc-pV6Z basis set, the higher-order connected quadruples contribution CCSDTQ–CCSDT(Q) using aug-cc-pVQZ, CCSDTQ5–CCSDTQ differences were obtained with the aug-cc-pVTZ basis set, and finally the very small CCSDTQ56–CCSDTQ5 term was obtained with the aug-cc-pVDZ basis set. CCSD(T)-level inner shell correlation corrections were obtained at the CCSD(T)/ACV{5,6}Z level, while post-CCSD(T) corrections to the core-valence contribution were obtained at the CCSDT(Q)/ACV5Z level. Scalar relativistic corrections were obtained at the CCSD(T)/aug-cc-pwCV5Z level using the 4$^{th}$-order Douglas-Kroll-Hess (DKH4) Hamiltonian[85–87] (see Refs.[88,89] for introductions to relativistic quantum chemistry), while the diagonal Born-Oppenheimer correction (DBOC) was obtained at the CCSD/aVQZ level using the implementation[90] in the CFOUR[91] program system.

**TABLE 3.** Calculated and experimental atomic electron affinities (eV) for 1$^{st}$ and 2$^{nd}$ row atoms

|    | CCSD$^a$ | post-CCSD$^b$ | CV$^c$ | Rel.$^d$ | SO$^e$ | DBOC$^f$ | Best Calc. | Expt.$^g$ | Uncertainty |
|----|----------|---------------|--------|----------|--------|----------|------------|-----------|-------------|
| H  | 0.75389  | 0             | 0      | -0.00004 | 0      | -0.00094 | 0.75290    | 0.754195[71]   | 0.000019    |
| B  | 0.18746  | 0.08907       | 0.00504 | -0.00123 | -0.0005 | 0.00011 | 0.27991    | 0.279723[73]   | 0.000025    |
| C  | 1.17297  | 0.08799       | 0.00761 | -0.00278 | -0.0037 | 0.00013 | 1.26226    | 1.262119[74]   | 0.000020    |
| O  | 1.28471  | 0.18405       | 0.00225 | -0.00611 | -0.0023 | 0.00012 | 1.46267    | 1.4611135[75,76] | 0.0000009   |
| F  | 3.24579  | 0.17690       | 0.00435 | -0.00963 | -0.0167 | 0.00009 | 3.40079    | 3.4011897[77]  | 0.0000024   |
| Al | 0.38645  | 0.07258       | -0.01631 | -0.00524 | -0.0036 | 0.00002 | 0.43389    | 0.43283[72]    | 0.00005     |
| Si | 1.35085  | 0.07548       | -0.00985 | -0.00779 | -0.0186 | 0.00001 | 1.39014    | 1.3895211[75]  | 0.0000013   |
| P  | 0.64860  | 0.09907       | -0.00412 | -0.00892 | 0.0111 | 0.00002  | 0.74577    | 0.746607[77]   | 0.000010    |
| S  | 1.99113  | 0.10347       | -0.00082 | -0.01173 | -0.0043 | 0.00001 | 2.07777    | 2.07710403[78] | 0.00000051  |
| Cl | 3.56195  | 0.09900       | 0.00129 | -0.01453 | -0.0365 | 0.00000 | 3.61124    | 3.612725[79]   | 0.000027    |

$^a$Computed at the CCSD(F12*)/d-awCV5Z level.
$^b$See Table 4 below for a further component breakdown.
$^c$i.e., a sum of CCSD(T)/awCV{5,6}Z, T$_3$-(T)/ACV5Z and (Q)/ACV5Z.
$^d$Scalar relativistic contribution from 4$^{th}$ order Douglas-Kroll method, calculated at the CCSD(T)/awCV5Z level.
$^e$From experimental fine structures[72,73,76–78,92,93].
$^f$Diagonal Born-Oppenheimer Corrections, obtained at the CCSD/AVQZ level.
$^g$Compilation in Ref.[94]

By inspection of Tables 3-4, important trends can be attributed for the various contributions to the calculated EAs. For instance, none of the post-(Q) contributions exceed ~2 meV, or ~0.2% of the total EA calculated. Thus, CCSDT(Q) may be treated as a fine approximation to the full CCSDTQ56 approached employed here (a similar conclusion is drawn in Ref[95]). As noted previously (e.g., Refs.[95,96]), the Core-Valence correlation contribution decreases in weight (i.e., percentage in the calculated EA) from left to right in the Periodic Table (see

Ref.[96]). Scalar relativistic corrections systematically reduce the electron affinity,[96–98] as they favor the more compact neutral atom over the more diffuse anion: they increase with atomic number from an essentially negligible 0.04 meV for hydrogen to quite nontrivial amounts of ~12 and ~14 meV for sulfur and chlorine, respectively. Spin-orbit (SO) corrections were not calculated, but taken from the experimental fine structures.[72,73,76–78,92,93] While the first-order splitting should scale roughly as $Z^2$ with the atomic number Z, the difference between neutral and anion exhibits somewhat more erratic behavior because of the different electronic states involved: notably, the closed-shell singlet F⁻ and Cl⁻, as well as systems with $^4$S ground states such as C⁻, Si⁻, and P, do not exhibit any first-order spin-orbit splitting.[96] The largest Diagonal Born-Oppenheimer correction (DBOC) applies to hydrogen, where it amounts to ~1 meV, while for the other elements it only ranges between ~0.1-0.01 meV. In all, it is possible to see that all of the best calculated values (found in the 'Calc.' column, summing up all of the calculated contributions) agree to ~1 meV with their experimental counterparts. Hence, we may argue that the calculated CCSD limit component should be accurate to ~1 meV as well.

**TABLE 4.** Post-CCSD corrections (eV) to the calculated atomic electron affinities

|    | (T)[a]  | T3-(T)[b] | (Q)[b]  | post-(Q)[c] | Total   |
|----|---------|-----------|---------|-------------|---------|
| H  | 0       | 0         | 0       | 0           | 0       |
| B  | 0.07000 | 0.01384   | 0.00450 | 0.00074     | 0.08907 |
| C  | 0.07598 | 0.00655   | 0.00557 | -0.00012    | 0.08799 |
| O  | 0.16840 | 0.00378   | 0.01477 | -0.00289    | 0.18405 |
| F  | 0.17919 | -0.01065  | 0.01121 | -0.00286    | 0.17690 |
| Al | 0.05780 | 0.01030   | 0.00363 | 0.00084     | 0.07258 |
| Si | 0.06577 | 0.00356   | 0.00522 | 0.00092     | 0.07548 |
| P  | 0.08477 | 0.00748   | 0.00617 | 0.00065     | 0.09907 |
| S  | 0.09563 | -0.00041  | 0.00763 | 0.00062     | 0.10347 |
| Cl | 0.10032 | -0.00932  | 0.00761 | 0.00038     | 0.09900 |

[a]Obtained using the ACV{5,6}Z extrapolation scheme.
[b]Obtained using the AV6Z basis set
[c]The aVQZ, aVTZ and aVDZ basis sets were used for the Q-(Q), $T_5$ and $T_6$ components, respectively.

As mentioned in the introduction, diffuse functions are needed to describe the behavior of electrons found relatively far from the nucleus in anionic systems. Thus, one should expect the augmented basis sets to be particularly useful for the calculation of EAs—in fact (*vide supra*), they were originally developed for this specific purpose by Dunning and coworkers.[20]

As a first test for the aVnZ-F12 basis sets, we will consider CCSD-F12b EAs of the first- and second-row atoms. In order to sidestep the issue of CCSD-F12b and CCSD allegedly converging to different basis set limits[39] (we have however shown in a previous paper[41] that most or all of this discrepancy can be removed through ensuring adequate radial flexibility in the orbital basis set), we are also obtaining our reference values at the CCSD-F12b level. Specifically, CCSD-F12b/d-aug-cc-pwCV5Z results were chosen as the reference, as d-aug-cc-pwCV5Z was found to be appropriate for obtaining the CCSD components of EAs (see Table 3 above), and is basically the largest orbital basis we can afford, given our hardware limitations (for obtaining our reference values, we used awCV5Z as the MP2fit, and aV5Z+d as the JKfit and OptRI auxiliary basis sets). Absolute errors for VnZ-F12 and aVnZ-F12 (n=D-5) with respect to the d-aug-cc-pwCV5Z reference values are presented in Table 5.

**TABLE 5.** Absolute errors (eV) for CCSD-F12b electron affinities obtained by the VnZ-F12 and the aVnZ-F12 basis set families. d-aug-cc-pwCV5Z is used as the reference.

| VnZ-F12 | | | | | VnZ-F12 | | | |
|---|---|---|---|---|---|---|---|---|
| n= | D | T | Q | 5 | n= | D | T | Q | 5 |
| H | 0.528 | 0.356 | 0.252 | 0.113 | Al | 0.064 | 0.025 | 0.020 | 0.006 |
| B | 0.063 | 0.039 | 0.019 | 0.012 | Si | 0.074 | 0.023 | 0.012 | 0.005 |
| C | 0.042 | 0.028 | 0.012 | 0.006 | P | 0.179 | 0.063 | 0.028 | 0.008 |
| O | 0.085 | 0.036 | 0.020 | 0.007 | S | 0.118 | 0.056 | 0.021 | 0.006 |
| F | 0.052 | 0.023 | 0.009 | 0.007 | Cl | 0.061 | 0.047 | 0.015 | 0.005 |
| aVnZ-F12 | | | | | aVnZ-F12 | | | |
| n= | D | T | Q | 5 | n= | D | T | Q | 5 |
| H | 0.013 | 0.011 | 0.006 | 0.003 | Al | 0.016 | 0.001 | 0.000 | 0.000 |
| B | 0.013 | 0.003 | 0.002 | 0.001 | Si | 0.026 | 0.001 | 0.001 | 0.000 |
| C | 0.013 | 0.000 | 0.000 | 0.000 | P | 0.122 | 0.024 | 0.005 | 0.002 |
| O | 0.064 | 0.016 | 0.004 | 0.001 | S | 0.074 | 0.024 | 0.002 | 0.001 |
| F | 0.035 | 0.011 | 0.002 | 0.001 | Cl | 0.022 | 0.027 | 0.002 | 0.000 |

For the first-row atoms, the VnZ-F12 sets perform very poorly. VDZ-F12 displays large errors for all elements, and in particular exhibits an error of no less than 0.528 eV for hydrogen. Choosing VTZ-F12 instead lowers the errors for hydrogen, boron and carbon by

~33%, and for oxygen and fluorine by ~67%. VQZ-F12 improves further, and cuts all VTZ-F12 errors by 50% or more, except for hydrogen, for which the error is only reduced by ~30%. V5Z-F12 performs even better, but still makes an error of 0.113 eV for hydrogen while reducing the errors for all other elements by a small amount (a few meV).

The aVnZ-F12 sets, on the other hand, clearly offer an advantage. aVDZ-F12 already cuts the error for hydrogen to only 13 meV, and even performs better than VTZ-F12 for boron and carbon. aVTZ-F12 cuts the error for hydrogen further to 11 meV, and is actually superior to VQZ-F12 for the other elements. aVQZ-F12 and aV5Z-F12 are pretty close in performance; both are superior to V5Z-F12 and produce results which are very close, if not equivalent, to the reference values.

Moving on to the second-row elements, a similar trend continues: aVDZ-F12 offers a significant improvement over VDZ-F12, as the former reduces the errors for aluminum, silicon and chlorine by more than 50%, and those for phosphorus and sulfur by more than 30%. For aluminum, silicon and phosphorus, aVTZ-F12 performs better than VQZ-F12, and still offers a significant improvement over VTZ-F12 for sulfur and chlorine. Again, aVQZ-F12 and aV5Z-F12 are superior to V5Z-F12 and produce results virtually equivalent to the reference values.

In all, it is clear that the aVnZ-F12 sets are useful for calculating EAs, and offer a noticeable improvement over their non-augmented counterparts.

**<u>Atomization energies</u>**

The W4-11 dataset was assembled by our group to include 140 total atomization energies of small first- and second-row molecules and radicals.[99] These cover a broad spectrum of bonding situations and multireference character, and as such may serve as a useful benchmark for parametrization and validation of various quantum chemistry methods (such as DFT functionals and composite protocols). We have recently revisited this dataset and adjusted it for the validation of the Wn-F12 (explicitly correlated) protocols,[41] thus removing beryllium-containing compounds and adding 60 more species to the list. The resulting dataset, containing 200 systems, is denoted W4-17.[100] In the present work, however, we were unable to complete the aV{Q,5}Z-F12 calculations for eight species due to near-linear dependence: these are cyclobutene, n-pentane, benzene, $C_2Cl_6$, borole, cyclopentadiene, pyrrole and cyclobutadiene.

Again, we decided to use F12b reference values; for considerations similar to those mentioned in the 'electron affinities' section above, CCSD-F12b/awCV5Z was chosen as the reference level.

TABLE 6. RMSDs (kcal/mol) for the SCF and CCSD-F12b components of the W4-17 dataset of total atomization energies. awCV5Z is used as the reference.

| | SCF | | | | | | |
|---|---|---|---|---|---|---|---|
| n | D | T | Q | 5 | | | |
| VnZ-F12 | 0.795 | 0.111 | 0.025 | 0.021 | | | |
| aVnZ-F12 | 0.683 | 0.105 | 0.025 | 0.021 | | | |
| | CCSD-F12b | | | | | | |
| n | D | T | Q | 5 | {D,T} | {T,Q} | {Q,5} |
| VnZ-F12 | 2.918 | 0.733 | 0.094 | 0.020 | 1.170 | 0.153 | 0.014 |
| aVnZ-F12 | 2.824 | 0.710 | 0.093 | 0.023 | 1.050 | 0.144 | 0.016 |

Examination of the error statistics for both SCF and CCSD-F12b TAE components (Table 6) reveals that the aVnZ-F12 basis sets do not offer a great improvement over their unaugmented counterparts: while it is true that among the double-zeta options, aVDZ-F12 outperforms VDZ-F12 by about 0.1 kcal/mol, aVnZ-F12 and VnZ-F12 sets display similar RMSDs from the reference data when it comes to the larger sets (n=T-5). Considering the fact that VDZ-F12 is in any case considered inadequate for molecular atomization energies (e.g., [41]), the slight advantage displayed by aVDZ-F12 is of no practical consequence.

The extrapolated V{T,Q}Z-F12 results agree slightly better with our reference values than aV{T,Q}Z-F12 — seemingly due to the somewhat inferior performance of VTZ-F12 — while neither basis family seems to be notably preferable for the {Q,5} extrapolation schemes. Thence, it is possible to deduce that the aVnZ-F12 sets are not particularly useful for atomization energies of neutral (and presumably cationic) species.

**Noncovalent Interactions Between Neutral Monomers**

**I. The S66 dataset**

The S66 dataset[101,102] was assembled by Hobza and coworkers to include 66 noncovalent dimers: the dimers were generated from 14 monomers in various combinations. The selection of monomers was based on their frequency as motifs or functional groups in the most commonly found biomolecules. The S66 set was designed with a balance in mind between electrostatics-dominated (hydrogen bonding), dispersion-dominated (including $\pi$ stacking), and mixed-influence complexes. Weak hydrogen bonds, aromatic–aliphatic, and aliphatic–aliphatic interactions are incorporated into the set as well, making it representative of NCIs one might see in biomolecules.

We shall use this dataset to discuss the basis set convergence for MP2-F12 dissociation energies, and examine the performance of the aVnZ-F12 basis set family compared to several alternatives.

**MP2-F12 basis set convergence**

The basis set error statistics for MP2-F12 dissociation energies of the S66 basis set, calculated with the AVnZ, sano-pVnZ, ano-pVnZ+, VnZ-F12 and aVnZ-F12 basis sets and different CABS for the S66 dataset, are presented in Table 7. CP-corrected MP2-F12/V{T,Q}Z-F12 results were used as the reference.

Let us start with the VDZ basis sets. Without CP corrections, conventional AVDZ performs poorly due to a considerable error in the HF+CABS component. The performance is improved, however, when a CP correction is employed. It is clear that sano-pVDZ is too small to provide useful correlation energies. When used "raw", ano-pVDZ+ performs better than the similarly-sized AVDZ, but applying CP corrections essentially cancels out this advantage. Among all "raw" results, the ones obtained by aVDZ-F12 are the most accurate, although no clear advantage over VDZ-F12 is observed (such an advantage does become a bit more apparent when a better CABS is used). Nonetheless, when CP corrections are used, aVDZ-F12 displays the best performance, and offers a noticeable improvement over the underlying VDZ-F12. A further improvement at negligible extra cost can be obtained by substituting the new[103] cc-pVDZ-F12/OptRI+ CABS for cc-pVDZ-OptRI: this reduces the error in the HF+CABS component as these OptRI+ basis sets were designed to do.

**TABLE 7.** RMS deviations (kcal/mol) of various basis sets and CABS for the S66 benchmark. MP2-F12/V{T,Q}Z-F12, full CP is used as the reference.

| Orbital Basis Set | CABS | raw | | CP | | half-CP | |
|---|---|---|---|---|---|---|---|
| | | HF+CABS | MP2-F12 | HF+CABS | MP2-F12 | HF+CABS | MP2-F12 |
| AVDZ | AVDZ/OptRI | 0.241 | 0.429 | 0.033 | 0.138 | 0.111 | 0.156 |
| sano-pVDZ | VDZ-F12/OptRI | 0.142 | 0.195 | 0.022 | 0.543 | 0.077 | 0.322 |
| ano-pVDZ+ | VDZ-F12/OptRI | 0.116 | 0.309 | 0.016 | 0.199 | 0.061 | 0.100 |
| VDZ-F12 | VDZ-F12/OptRI | 0.075 | 0.096 | 0.010 | 0.150 | 0.034 | 0.038 |
| aVDZ-F12 | VDZ-F12/OptRI | 0.069 | 0.097 | 0.010 | 0.085 | 0.032 | 0.029 |
| aVDZ-F12 | VDZ-F12/OptRI+ & diffuse | 0.049 | 0.078 | 0.004 | 0.090 | 0.024 | 0.028 |
| AVTZ | AVTZ/OptRI | 0.044 | 0.163 | 0.007 | 0.031 | 0.020 | 0.068 |
| sano-pVTZ | VTZ-F12/OptRI | 0.039 | 0.115 | 0.008 | 0.177 | 0.022 | 0.109 |
| ano-pVTZ+ | VTZ-F12/OptRI | 0.028 | 0.088 | 0.005 | 0.099 | 0.015 | 0.059 |
| VTZ-F12 | VTZ-F12/OptRI | 0.030 | 0.069 | 0.002 | 0.054 | 0.014 | 0.015 |
| aVTZ-F12 | VTZ-F12/OptRI | 0.019 | 0.045 | 0.002 | 0.019 | 0.009 | 0.014 |
| AVQZ | AVQZ/OptRI | 0.019 | 0.073 | 0.002 | 0.015 | 0.009 | 0.030 |
| sano-pVQZ | VQZ-F12/OptRI | 0.028 | 0.063 | 0.003 | 0.073 | 0.015 | 0.045 |
| ano-pVQZ+ | VQZ-F12/OptRI | 0.024 | 0.056 | 0.003 | 0.050 | 0.013 | 0.031 |
| VQZ-F12 | VQZ-F12/OptRI | 0.009 | 0.029 | 0.000 | 0.015 | 0.005 | 0.008 |
| aVQZ-F12 | VQZ-F12/OptRI & diffuse | 0.003 | 0.007 | 0.000 | 0.008 | 0.001 | 0.004 |
| AV5Z | AV5Z/OptRI | 0.002 | 0.019 | - | - | - | - |
| V5Z-F12 | AVQZ/MP2FIT | 0.001 | 0.009 | 0.000 | 0.007 | 0.001 | 0.005 |

Moving on to the VTZ sets, sano-pVTZ performs relatively poorly, while ano-pVTZ+ offers a general improvement (with and without CP corrections) over AVTZ. The latter performs badly without CP corrections, but actually beats VTZ-F12 but when such are applied. aVTZ-F12, however, is now clearly superior over VTZ-F12 as well as the other options.

As for the raw VQZ sets, aVQZ-F12 results are virtually equivalent in quality to their costlier V5Z-F12 counterparts (It should be noted that we had to remove the diffuse $f$ function on carbon to avoid near-linear dependence issues with the benzene complexes.). With CP, the same is also true of results obtained by VQZ-F12. sano-pVQZ actually performs worse than aVTZ-F12, while ano-pVQZ+ may be slightly preferable over AVQZ when used "raw".

Finally, we find V5Z-F12 — and hence, by extension, aVQZ-F12 — to be remarkably superior to AV5Z for the S66 dataset.

**II. BEGDB water clusters subset – n-body decomposition**

The water clusters benchmark in Hobza's BEGDB (Benchmark Energy and Geometry Data Base)[104] covers different isomers of neutral water clusters from $(H_2O)_2$ to $(H_2O)_{10}$; their geometries and energetics were collected from earlier works by Shields and coworkers[105]. It often serves for the purpose of training and validation of approximate methods, such as DFT functionals.

Decomposition of the dissociation energies of the water clusters into n-body contributions affords insights into many-body interactions[106–109] in water clusters, which is essential for developing accurate analytical representations of the water potential energy surface.[110,111] We also showed in a very recent benchmark study,[46] using conventional and explicitly correlated coupled cluster methods, that very close agreement with very large basis set whole-cluster calculations can be achieved as long as basis set convergence has been attained in the 3-body and especially 2-body terms. Hence, it becomes possible to accurately treat larger clusters such as isomers of $(H_2O)_{20}$ using post-MP2 corrections for just 2- and 3-body terms.[46,112]

In a recent study conducted by our group,[46] we found that four-body and higher terms are recovered fairly well at the MP2-F12/cc-pVTZ-F12 level for water clusters $(H_2O)_2$ through $(H_2O)_6$ of the BEGDB water clusters subset. In contrast, the three-body terms and especially two-body terms were found to be sensitive to both basis set and electron correlation level. Thus, the 2-body term was captured through a CCSD(F12*)/cc-pV5Z-F12 + [CCSD(T)−CCSD]/haV5Z composite scheme, while the 3-body term was obtained using CCSD(F12*)/cc-pVQZ-F12 + [CCSD(T)−CCSD]/haVQZ.

In the present work, we have analyzed the CCSD 2-body and 3-body contributions to the cohesive energies of $(H_2O)_2$ through $(H_2O)_6$. For the reference level, we had to choose a sufficiently large basis set and yet avoid numerical complications. Thus, we debated whether to choose haV{5,6}Z, V{Q,5}Z-F12, or, alternatively, to use their average $\frac{\text{haV\{5,6\}Z} + \text{V\{Q,5\}Z} - \text{F12}}{2}$. In order to decide between them, we calculated the raw, CP and half-CP corrected cohesive energy for the water dimer (being a model system for a 2-body contribution to the cohesive energy of any n-mer) using all three options, and found out that all calculated values are found within the range of 4.815–4.824 kcal/mol. Attempted calculations using even larger basis sets such as CCSD-F12b/awCV5Z proved fruitless owing to near-linear dependence problems. Considering that: (a) all three choices (namely, CCSD/haV{5,6}, CCSD(F12*)/V{Q,5}Z-F12, and the average of the two) yield very similar results; (b) the raw and counterpoise-corrected extrapolated limits are essentially identical; (c) a truncated n-body analysis without counterpoise is computationally much more efficient; we finally decided on the average $\frac{\text{haV\{5,6\}Z} + \text{V\{Q,5\}Z} - \text{F12}}{2}$ without CP correction as our chosen reference level for the 2-body terms. Full results are given in the ESI, while RMSDs for smaller basis sets and extrapolations therefrom are presented in Table 8.

Let us first consider the 2-body term. When it comes to the conventional CCSD/haVnZ calculations, haVTZ and haVQZ perform rather poorly, with RMSDs of 0.030 and 0.025 kcal/mol, respectively. RMSDs are substantially reduced when haV5Z (0.004 kcal/mol) and haV6Z (0.003 kcal/mol) are used; however, such large basis sets might be impractical due to the heavy computational cost they entail. The haV{Q,5}Z extrapolation does not offer a significant improvement over haVQZ (0.019 kcal/mol RMSD), and actually performs worse than haV5Z alone.

**TABLE 8.** RMS deviations (kcal/mol) of various basis sets for the CCSD 2- and 3-body contributions to the cohesive energies of $(H_2O)_2$ through $(H_2O)_6$. $\frac{haV\{5,6\}Z+V\{Q,5\}Z-F12}{2}$ is used as the reference for the 2-body term, while aV{T,Q}Z-F12 is used as such for the 3-body term. The CCSD(F12*) approximation was used for all explicitly correlated calculations.

| Level of theory | 2-body | 3-body |
|---|---|---|
| CCSD/haVTZ | 0.030 | 0.014 |
| CCSD/haVQZ | 0.025 | 0.007 |
| CCSD/haV5Z | 0.004 | 0.004 |
| CCSD/haV6Z | 0.003 | - |
| CCSD/haV{Q,5}Z | 0.019 | 0.004 |
| CCSD(F12*)/VTZ-F12 | 0.038 | 0.004 |
| CCSD(F12*)/VQZ-F12 | 0.013 | 0.003 |
| CCSD(F12*)/V5Z-F12 | 0.009 | - |
| CCSD(F12*)/V{T,Q}Z-F12 | 0.004 | 0.003 |
| CCSD(F12*)/aVTZ-F12 | 0.020 | 0.005 |
| CCSD(F12*)/aVQZ-F12 | 0.004 | 0.001 |
| CCSD(F12*)/aV{T,Q}Z-F12 | 0.002 | REF |
| (haV{5,6}Z+V{Q,5}Z-F12)/2 | REF | - |

Now let us consider CCSD(F12*) calculations with the VnZ-F12 and aVnZ-F12 basis sets. VTZ-F12 actually has a larger RMSD (0.038 kcal/mol) than haVTZ. VQZ-F12, on the other hand, performs better than haVQZ with an RMSD of 0.013 kcal/mol. V5Z-F12 offers a minor improvement (0.009 kcal/mol RMSD), while V{T,Q}Z-F12 actually performs comparably to haV5Z and haV6Z at a substantially reduced computational cost.

What about the aVnZ-F12 basis sets? aVTZ-F12 displays the best performance among the triple-zeta basis sets at hand, with an RMSD of 0.020 kcal/mol. Similarly, the aVQZ-F12 displays superior performance relative to haVQZ and VQZ-F12, with an RMSD comparable to that obtained by haV5Z and V{T,Q}Z-F12. Finally, the aV{T,Q}Z-F12 extrapolations offer the lowest RMSD, 0.002 kcal/mol. Thus, it even performs better than V5Z-F12, and indeed the even larger haV6Z basis set, yielding results that are basically indistinguishable from the reference values.

Let us move on to the 3-body term. Due to the superior performance of the computationally somewhat affordable aV{T,Q}Z-F12 for the 2-body term, we decided to use it as the reference for the 3-body results. Despite the fact that the 3-body contribution to the energy is much smaller than that of the 2-body term (and thus assessing the performance of the different basis sets becomes harder, since fewer significant digits are available), similar trends can be seen. Again, haVTZ and haVQZ do not perform very well, with RMSDs of 0.014 and 0.007 kcal/mol, respectively. haV5Z and haV{Q,5}Z do not offer a significant improvement over the much smaller VTZ-F12 and aVTZ-F12 — and even though VQZ-F12 and V{T,Q}Z-F12 perform quite well, both are outperformed by aVQZ-F12, which produces results very close to the reference values.

Summing up: for the neutral NCIs calculated for both S66 dataset and BEGDB water clusters, the aVnZ-F12 basis set family appears to be superior, or at least as good as the available alternatives.

**Neutral + Anionic Noncovalent Interactions**

The WATER27 dataset, a component of the GMTKN24[113] and GMTKN30[114] benchmark suites, was first introduced by Bryantsev *et al.*:[115] it consists of ten neutral structures of $(H_2O)_n$ (n=2-8), four isomers of $(H_2O)_{20}$, five protonated water clusters $H_3O^+(H_2O)_n$ (n=1-3, 6), seven hydrated hydroxide clusters $OH^-(H_2O)_n$ (n=1-6), and one hydroxonium-hydroxide zwitterion, $H_3O^+(H_2O)_6OH^-$. All geometries were obtained at the B3LYP/6-311++G(2d,2p) level of theory in the original reference and used here unmodified. The original benchmark calculations included MP2/CBS energies, corrected with CCSD(T)/aug-cc-pVDZ high level corrections (HLC) for all except the four $(H_2O)_{20}$ isomers. These four structures were found to be too large for a convergence study of their CCSD energies: hence, we used the remaining subset of 23 structures (denoted WATER23 from now on) for our study.

For obtaining CCSD(F12*) cohesive energies, we managed to run calculations up to VTZ-F12 and aVTZ-F12 for the whole WATER23 dataset. However, based upon our recent inspection of this dataset, we chose MP2-F12/aV{T,Q}Z-F12, corrected with a CCSD/aV{D,T}Z-F12 HLC, as our reference level. Counterpoise corrections at this level of theory are quite small (all are found within 0.014 kcal/mol RMSD from each other, see Table 9) and hence indicative of a regime where full counterpoise should be closest to the basis set limit.[66]

In order to confirm that basis set convergence has indeed been achieved, we also calculated VQZ-F12 and aVQZ-F12 values for $OH^-(H_2O)_n$ (n=1-4) clusters. As expected, for the latter clusters, MP2-F12/aV{T,Q}Z-F12 + HLC exhibits excellent performance, with RMSDs of 0.020 (raw), 0.013 (half-CP), and 0.008 (full-CP) relative to the CP corrected aV{T,Q}Z-F12 results. In addition, it is possible to see that full-CP corrected MP2-F12/aV{T,Q}Z-F12 + HLC performs well for the $OH^-(H_2O)_n$ (n=1-4) clusters as well as the neutral and protonated clusters.

We now return to the error statistics of WATER23. For its anionic subset, we find clearly unacceptable "raw" errors with VnZ-F12 (RMSD=0.4 kcal/mol for VDZ-F12, and still 0.19 kcal/mol for VQZ-F12), which are drastically reduced for aVnZ-F12 (to 0.13 and 0.04 kcal mol, respectively). For "raw" calculations on the remaining neutral and cationic species, aVnZ-F12 still offers a substantial improvement in RMSD (from 0.13 to 0.05 for n=D, from 0.08 to 0.04 for n=T). It is true that counterpoise correction followed by extrapolation suppresses most of the RMSD for VnZ-F12, with full CP V{D,T}Z-F12 clocking in at just 0.013 kcal/mol for the whole dataset. However, the need for counterpoise correction becomes a significant computational obstacle when treating a large cluster, e.g. $(H_2O)_{20}$, by a truncated n-body expansion. Besides, one can argue that it might be hazardous to rely on what is clearly another fortuitous error cancellation.

Table 9: RMS Deviations (kcal/mol) for the CCSD(F12*) cohesive energies calculated with various basis sets for the WATER23 dataset. MP2-F12/aV{T,Q}Z-F12, corrected with a CCSD/aV{D,T}Z-F12 HLC, is chosen as the reference level.

|  | WATER23 | | | Only neutral and protonated water clusters | | | Only deprotonated water clusters | | |
| --- | --- | --- | --- | --- | --- | --- | --- | --- | --- |
|  | raw | CP | half | raw | CP | half | raw | CP | half |
| VDZ-F12 | 0.242 | 0.222 | 0.067 | 0.131 | 0.217 | 0.044 | 0.402 | 0.177 | 0.123 |
| VTZ-F12 | 0.173 | 0.069 | 0.062 | 0.081 | 0.068 | 0.007 | 0.294 | 0.054 | 0.123 |
| VQZ-F12 |  |  |  |  |  |  | 0.189 | 0.015 | 0.088 |
| V{D,T}Z-F12 | 0.185 | 0.013 | 0.094 | 0.090 | 0.010 | 0.041 | 0.308 | 0.018 | 0.162 |
| V{T,Q}Z-F12 |  |  |  |  |  |  | 0.163 | 0.002 | 0.082 |
| aVDZ-F12 | 0.081 | 0.195 | 0.065 | 0.053 | 0.190 | 0.070 | 0.132 | 0.163 | 0.027 |
| aVTZ-F12 | 0.072 | 0.043 | 0.024 | 0.042 | 0.046 | 0.004 | 0.117 | 0.020 | 0.051 |
| aVQZ-F12 |  |  |  |  |  |  | 0.039 | 0.007 | 0.016 |
| aV{D,T}Z-F12 | 0.094 | 0.024 | 0.058 | 0.062 | 0.009 | 0.035 | 0.138 | 0.045 | 0.092 |
| aV{T,Q}Z-F12 |  |  |  |  |  |  | 0.021 | REF | 0.010 |
| MP2-F12/aV{T,Q}Z-F12 + HLC | 0.014 | REF | 0.007 | 0.012 | REF | 0.006 | 0.020 | 0.008 | 0.013 |

In contrast, aV{T,Q}Z-F12 can get down to 0.02 kcal/mol without any counterpoise correction, and is hence more suitable for 2-body and 3-body terms of large cluster (say, (H$_2$O)$_{20}$ or larger): we note that even for aVQZ-F12 without extrapolation, the RMSD is already down to 0.04 kcal/mol.

Considering only neutral and protonated clusters, VDZ-F12 at first sight seems to be superior to aVDZ-F12, and is surprisingly close to the reference values. That being said, the fact that raw and full-CP corrected VDZ-F12 results exhibit much inferior performance suggests that the half-CP ones benefit from a fortuitous error cancellation between basis set superposition error and what was termed "intrinsic basis set insufficiency" in Ref.[66]. Full-CP corrected V{D,T}Z-F12 also performs exceptionally well, with only 0.013 kcal/mol RMSD for the whole dataset, and 0.010 kcal/mol RMSD when only neutral and protonated water clusters are considered. However, we recommend treating this seemingly successful performance with caution, due to the inferior results for each of the basis sets used in the extrapolated scheme under consideration (namely, VDZ-F12 and VTZ-F12). These latter results might indeed indicate that yet another fortuitous error cancelation is taking place here. Finally, full-CP corrected aV{D,T}Z-F12 displays satisfactory performance, while no evidence for contingent error cancelations are at hand.

In conclusion, when it comes to the WATER23 dataset, full-CP corrected aV{T,Q}Z-F12 performs best for the MP2-F12 correlation component, and full-CP corrected aV{D,T}Z-F12 does so for the [CCSD(F12*) – MP2-F12] HLC.

**Summary and Conclusions**

We have developed and benchmarked a new family of basis sets, augmented with additional diffuse functions, for explicitly correlated calculations. These sets, denoted aug-cc-pVnZ-F12 (or aVnZ-F12 for short), are now available for the atoms H, B-Ne, and Al-Ar. Considering their performance for electron affinities of first- and second-row elements, the W4-17 dataset of total atomization energies, the S66 dataset of noncovalent interactions, the BEGDB water clusters subset, and the WATER23 subset of the GMTKN24[113] and GMTKN30[114] benchmark suites, we can draw the following conclusions:

- In the context of calculating accurate electron affinities, the aVnZ-F12 (n=D-5) sets perform systematically better than the corresponding VnZ-F12 ones: Convergence to the reference values is already achieved using aVQZ-F12, while even V5Z-F12 presents an error of no less than 0.1 kcal/mol for hydrogen and 0.01 for boron. Similarly, the errors made by VQZ-F12 are an order of magnitude larger than those of aVQZ-F12.

- For the calculation of total atomization energies of neutral molecules, the larger aVnZ-F12 basis sets offer no significant advantage over VnZ-F12.

- In contrast, for noncovalent interactions between neutral species, the aVnZ-F12 sets do outperform the VnZ-F12 ones, as well as other contenders such as AVnZ, sano-pVnZ, and ano-pVnZ+.

    I. For the diverse S66 dataset: Both "raw" and CP-corrected aVQZ-F12 results were found to be equivalent to those of V5Z-F12. Thus, the aVQZ-F12 basis was marked as a useful compromise between accuracy and computational cost for this dataset.

    II. For the 2- and 3-body contributions to the cohesive energies of the BEGDB water clusters subset, aVQZ-F12 and aV{T,Q}Z-F12 performed comparably to basis sets as large as haV5Z, V5Z-F12 and even haV6Z (available only for the 2-body term).

- As already reported previously,[46] for the anionic, deprotonated water clusters among the WATER23 benchmark, basis set convergence along the VnZ-F12 sequence is erratic, while aVnZ-F12 exhibits smooth convergence. Full-CP corrected aV{T,Q}Z-F12 seems to be the ultimate choice for the MP2-F12 correlation component, while aV{D,T}Z-F12 is recommended for a CCSD(F12*)–MP2-F12 HLC.

- We have also learned that complementary-auxiliary-basis-sets (CABS), used in explicitly correlated electronic structure calculations, are more transferable than generally assumed (e.g., AVnZ/OptRI produces reasonable results when used for ano-pVnZ+). That being said, OptRI+ does offer a noticeable improvement for VDZ-F12, and also for aVnZ-F12 (n=T,Q) when armed with an extra layer of diffuse functions (see 'Appendix: the choice of CABS for explicitly correlated calculations').

**Acknowledgements**

NS acknowledges a Feinberg Graduate School graduate fellowship. This research was supported by the Israel Science Foundation (grant 1358/15), the Minerva Foundation, and the Helen and Martin Kimmel Center for Molecular Design (Weizmann Institute of Science).

**Supporting Information**

Electronic supporting information (ESI), which includes charts describing the optimized augmenting functions in the aVnZ-F12 (n=D-5) basis sets, and spreadsheets containing EAs for 1$^{st}$ and 2$^{nd}$ row atoms, W4-17 atomization energies, S66 dissociation and cohesive energies for different basis sets and CABS combinations, n-body contributions for dissociation and cohesive energies of BEGDB water clusters, and CCSD-level dissociation and cohesive energies for the WATER23 subset, is available free of charge via the Internet at http://xxx. In addition, aVnZ-F12 (n=D-5) basis set files, CABS files, and an input example for a DF-MP2-F12/aVDZ-F12 calculation on $(H_2O)_2$, using VDZ-F12/OptRI+ with extra diffuse functions, are also attached (all in MOLPRO format).

**Appendix: The choice of CABS for explicitly correlated calculations**

For the two aug-cc-pVnZ and VnZ-F12 (n=D,T,Q,5) families, compact aug-cc-pVnZ/OptRI and VnZ-F12/OptRI CABS basis sets have been optimized.[116,117] Very recently, Shaw and Hill published improved VnZ-F12/OptRI+ basis sets,[103] in which a few extra functions, optimized for the CABS-corrected HF energy, are added. For the largest F12-optimized orbital basis set, V5Z-F12, no OptRI basis set is currently available: the original V5Z-F12 paper[45] recommends using the aug-cc-pwCV5Z/MP2fit basis set as an *ad hoc* choice.

**TABLE 10.** RMS deviations (kcal/mol) of various basis sets and CABS for the S66 benchmark. For each orbital basis set and CABS combination, the same orbital basis, combined with the aug-cc-pwCV5Z/MP2fit CABS, is used as the reference.

| Orbital Basis Set | CABS | $N_{CABS}^a$ | raw | | CP | |
|---|---|---|---|---|---|---|
| | | | *HF+CABS* | *E2corr* | *HF+CABS* | *E2corr* |
| ano-pVDZ+ | VDZ-F12/OptRI | 1056 | 0.043 | 0.131 | 0.003 | 0.038 |
| ano-pVDZ+ | AVDZ/OptRI | 1092 | 0.016 | 0.033 | 0.002 | 0.040 |
| ano-pVTZ+ | VTZ-F12/OptRI | 1392 | 0.006 | 0.003 | 0.001 | 0.005 |
| ano-pVTZ+ | AVTZ/OptRI | 1284 | 0.005 | 0.004 | 0.001 | 0.005 |
| ano-pVQZ+ | VQZ-F12/OptRI | 1632 | 0.008 | 0.004 | 0.001 | 0.004 |
| ano-pVQZ+ | AVQZ/OptRI | 1824 | 0.002 | 0.006 | 0.000 | 0.003 |
| VDZ-F12 | VDZ-F12/OptRI | 1056 | 0.029 | 0.010 | 0.012 | 0.016 |
| VDZ-F12 | VDZ-F12/OptRI+ | 1164 | 0.012 | 0.012 | 0.005 | 0.015 |
| VTZ-F12 | VTZ-F12/OptRI | 1392 | 0.017 | 0.002 | 0.003 | 0.002 |
| VTZ-F12 | VTZ-F12/OptRI+ | 1500 | 0.024 | 0.002 | 0.001 | 0.002 |
| VQZ-F12[b] | VQZ-F12/OptRI | 1632 | 0.006 | 0.001 | 0.000 | 0.000 |
| aVDZ-F12 | VDZ-F12/OptRI | 1056 | 0.029 | 0.011 | 0.012 | 0.014 |
| aVDZ-F12 | VDZ-F12/OptRI+ | 1164 | 0.010 | 0.013 | 0.006 | 0.013 |
| aVDZ-F12 | VDZ-F12/OptRI+ & diffuse | 1464 | 0.008 | 0.008 | 0.005 | 0.003 |
| aVTZ-F12 | VTZ-F12/OptRI | 1284 | 0.010 | 0.003 | 0.003 | 0.001 |
| aVTZ-F12 | VTZ-F12/OptRI+ | 1500 | 0.015 | 0.004 | 0.001 | 0.001 |
| aVTZ-F12 | VTZ-F12/OptRI+ & diffuse | 1692 | 0.006 | 0.004 | 0.001 | 0.000 |
| aVQZ-F12 | VQZ-F12/OptRI | 1632 | 0.002 | 0.001 | 0.000 | 0.001 |
| aVQZ-F12 | VQZ-F12/OptRI & diffuse | 1932 | 0.001 | 0.001 | 0.000 | 0.001 |

(a) number of CABS functions for benzene dimer, by way of illustration.
(b) Using VQZ-F12/OptRI+ instead of VQZ-F12/OptRI produces virtually indistinguishable results.

In the course of the current study, we re-examined the S66 benchmark for noncovalent interactions using the ano-pVnZ+ basis sets of Valeev and Neese,[63] for which no optimized

CABS are available. This led us to wonder about the transferability of OptRI basis sets between families. As a result, we decided to assess the performance of various orbital basis sets and CABS combinations; for this purpose, each orbital basis set was combined with aug-cc-pwCV5Z/MP2fit as CABS to serve as a reference level for smaller CABS choices combined with the same orbital basis. The resulting statistics are given in Table 10.

When combined with ano-pVDZ+, VDZ-F12/OptRI exhibits large RMSDs – particularly in the absence of CP corrections. AVDZ/OptRI performs much better, and is actually more precise than VDZ-F12/OptRI for VDZ-F12 itself.

For ano-pVTZ+, VTZ-F12/OptRI and AVTZ/OptRI offer comparable performance, and both are outperformed by VTZ-F12/OptRI for VTZ-F12. Similarly, there is little to choose between AVQZ/OptRI and VQZ-F12/OptRI when combined with ano-pVQZ+.

What about the recently proposed VnZ-F12/OptRI+ basis sets? These were designed to improve the HF+CABS part without sacrificing the MP2-F12 correlation energy. When applied to VDZ-F12, the added functions do improve the HF+CABS part with reference to ordinary VnZ-F12/OptRI. The MP2-F12 correlation energy, however, does not display such an improvement. In a similar manner, VTZ-F12/OptRI+ improves the HF+CABS part a bit for VTZ-F12 (only with CP), but no measurable effect of the additional functions on the correlation energy is seen. Finally, for VQZ-F12, the effect of the additional functions in OptRI+ is, at most, 0.001 kcal/mol for the S66 benchmark.

Let us turn our attention to the aVnZ-F12 orbital basis sets. It can be seen that adding one layer of diffuse functions to the VDZ-F12/OptRI+ CABS basis set produces an ultimate choice for aVDZ-F12, as the extra diffuse functions appear to be helpful for the correlation energy, while OptRI+ improves the HF+CABS part (The exponents for the diffuse layer were obtained by "even-tempered" extrapolation from the two outermost exponents). The same is true for aVTZ-F12 (in this case, however, the improvement is less substantial), while for aVQZ-F12, the original VQZ-F12/OptRI is also adequate.

Despite all of the above, we still learn that CABS are more transferable than generally assumed, at least between basis sets of similar size: for example, AVnZ/OptRI works acceptably well for ano-pVnZ+ and, by extension, to sano-pVnZ. For the S66 dataset, we see that OptRI+ offers a significant improvement over OptRI for VDZ-F12, yet represents no significant improvement for the rest of the VnZ-F12 orbital basis sets (n=T,Q). Such considerations served us for choosing an appropriate CABS for the orbital basis sets under consideration.

(A preliminary version of this Appendix was previously presented at the 2017 International Conference on Computational Materials Science and Engineering, ICCMSE-2017, in Thessaloniki, Greece, and made available as a preprint.[118])